\documentclass[aps,pra,twocolumn,showpacs,superscriptaddress]{revtex4}  
\usepackage{graphicx}  
\usepackage{dcolumn}   
\usepackage{bm}        
\usepackage{amssymb}   
\usepackage{verbatim}  
\usepackage{color}

\newcommand{\abs}[1]{\ensuremath{\left| #1 \right|}}

\newcommand{\beq}{\begin{equation}}
\newcommand{\eeq}{\end{equation}}

\newcommand{\eq}[1]{{(\ref{#1})}}
\newcommand{\commentout}[1]{{}}
\newcommand{\idx}{{l}}

\definecolor{red}{rgb}{1,0,0}
\newcommand{\colr}[1]{{#1}}
\begin{document}



\title{Quantum dynamics of instability-induced
pulsations of a Bose-Einstein condensate in an optical lattice}
\author{Uttam Shrestha}
\affiliation{Department of Physics, University of Connecticut,
Storrs, CT 06269-3046}
\author{Juha Javanainen}
\affiliation{Department of Physics, University of Connecticut,
Storrs, CT 06269-3046}
\author{Janne Ruostekoski}
\affiliation{School of Mathematics, University of Southampton,
Southampton, SO17 1BJ, UK}
\date{\today}

\begin{abstract}
We study the dynamics of a Bose-Einstein condensate in a one-dimensional optical lattice in the limit of weak atom-atom interactions, including an approximate model for quantum fluctuations. A pulsating dynamical instability in which atoms periodically collect together and subsequently disperse back into the initial homogeneous state may occur in the time evolution. We take into account the quantum fluctuations within the truncated Wigner approximation. We observe that the quasiperiodic behavior still persists for a single realization that represents a typical experimental outcome, but ensemble averages show various manifestations of quantum fluctuations. Quantum effect\colr{s} become more prominent \colr{when the effective interaction strength is increased}.

\end{abstract}

\maketitle

\section{Introduction}

The superfluidity of a Bose-Einstein condensate (BEC) in an optical lattice has been drawing considerable attention in the last several years \cite{Morsch}. As is well known, superflow of the BEC in free space suffers from an instability when the flow velocity reaches a critical value. Such an instability, known as Landau or energetic instability, exists when the superfluid flow is not a local minimum of energy and the system may lower its energy by emitting phonons \cite{Pethick}. In an optical lattice, in addition to energetic instability, the BEC may also exhibit dynamical or modulational instabilities that have been the subject of much experimental and theoretical research over recent years \cite{BUR01,Wu,SME02,CAT03,CRI04,FAL04,SAR05,WU03, Trombettoni,FER05,ZHE04,MON04,RUO05,BAN06,RUO07,Barontini,Ferris,Shrestha}. When the system is in the dynamically unstable regime, small perturbations grow exponentially in time resulting in irregular dynamics, loss of coherence, or an abrupt stop of the transport of the atom cloud \cite{Wu, SME02}.

In this paper we study dynamical instabilities of atoms in an optical lattice for the case of weak atom-atom interactions and also taking into account quantum fluctuations of the atoms. We recently reported~\cite{Shrestha} that, by appropriately selecting the strength of the atom-atom interactions, the corresponding classical system may exhibit a pulsating dynamical instability in which the atoms nearly periodically collect to a peak in lattice occupation numbers, and subsequently disperse back to (very close to) the initial unstable state. This is different from the conventional view, valid at strong interatomic interactions, that dynamical instabilities for BECs in optical lattices are associated with irregular dynamics.  \colr{The dynamics of an integrable double-well system provides a qualitative explanation of the pulsating phenomenon~\cite{Shrestha}:} Although the instability is a result of the interplay between the lattice discreteness and the nonlinearity that makes the lattice non-integrable, the dynamics of the lattice with many sites approximates the dynamics of an integrable system. Related classical pulsations starting from already compressed atom distribution in a lattice have been discussed in \cite{Barontini} within the framework of the nonpolynomial Schr\"odinger equation.

\colr{As the emergence of dynamical instabilities can also be closely related to various fundamental effects in nonequilibrium quantum dynamics, for example dissipative transport \cite{RUO05} and quantum phase transitions \cite{POL05}, it is particularly interesting to address how an entirely classical description of the pulsating instability~\cite{Shrestha}, embodied in the Discrete Nonlinear Schr\"odinger equation (DNLSE), is modified as a result of quantum fluctuations. In this paper we provide an approximate analysis of the quantum effects on the pulsating instability, as well as amend the discussion of Ref.~\cite{Shrestha} with additional details and angles.}


We consider different dynamically stable initial stationary superfluid flow states that are instantaneously transferred to a dynamically unstable regime by changing the strength of atom-atom interactions. We incorporate quantum fluctuations of the atoms using a stochastic phase-space method, the Truncated Wigner Approximation (TWA),~\colr{\cite{DRU93,Steel,SIN02,GAR02,ISE06}}. We find that the quasiperiodic behavior is still observable in individual stochastic realizations that represent typical outcomes of individual experiments. In fact, the damping of the quasiperiodic pulsating instability in a single stochastic trajectory remains very weak even in the presence of enhanced quantum fluctuations and for the case of an initial state with a substantial noncondensate atom fraction. However, as the timing, shape and location of the pulsation events in each realization change due to quantum effects, in the quantum mechanical ensemble average the wave function revival become progressively weaker, with an increased damping rate, when the effective interaction strength is increased. Other ensemble averages provide additional quantitative information about the effects of quantum fluctuations. For instance, in the limit of weak quantum fluctuations, the amplitude of the pulsating instability approaches the classical value obtainable from the DNSLE with a weak random noise seeding the instability, but for enhanced quantum fluctuations both the amplitude and the amplitude fluctuations of the pulsations become significantly larger than in the classical system.

In Sec.~\ref{TM} we formulate the theoretical model, mainly the Gross-Pitaevskii equation (GPE) \cite {Gross} and its discrete variant, the DNLSE \cite{SME02, Christ}. We use linear stability analysis to find the region of interaction strengths and flow quasimomenta  where the system develops an instability. In Sec.~\ref{TE} we investigate the time evolution of the DNLSE,  i.e., classical mean-field theory. Although the system initially develops an instability, the subsequent time evolution shows nearly periodic recurrence to the initial state. In Sec.~\ref{DW} we review the well understood double well system and argue that the dynamical behavior of the multi-site system is analogous to the two-site system, at least in the limit of weak nonlinearity. In Sec.~\ref{TWA} we study the dynamics beyond the classical mean field theory using the TWA. We compare various physical properties such as the number fluctuations and the overlaps of the state of the system in  single realizations with an ensemble averages in a few quantitative examples. The basic phenomenon of pulsations survive quantum corrections, but substantial modifications to the classical picture emerge when the atom number is \colr{reduced while keeping the chemical potential fixed}.

\section{\label{TM}Theoretical Model:  DNLSE}

At zero temperature the dynamics of a \colr{weakly interacting} BEC in an optical lattice can be modeled by the mean field Gross-Pitaevskii equation \cite{Gross,Pethick}

\begin{eqnarray}
i\hbar {d\Psi\over dt}
=\left(-{\hbar^2\over 2 m}{\Delta}+V({\bf x})+g_{3}|\Psi|^2 \right ) \Psi,
\label{ee1}
\end{eqnarray}
where $\Psi({\bf x},t)$ is a wave function corresponding to the bosonic field operator such that $|\Psi({\bf x})|^2 =n({\bf x})$ equals the atom density.
The coupling constant $g_{3}$  is related to the scattering length through $g_{3}={4\pi \hbar^2 a}/{m}$, where $a$ and $m$ are the $s$-wave scattering length and atomic mass. The positive and negative  scattering lengths respectively correspond to the repulsive and attractive atom-atom interactions. The Eq.~(\ref {ee1}) is an approximate description of an assembly of a large number of bosonic atoms that are in the same quantum mechanical state. Since the equation is nonlinear, the coefficient $g_{3}$ depends on the normalization of the wave function. Here the wave function is normalized to atom number $N$, so that we have
\begin{equation}
\int d^3x\,|\Psi({\bf x})|^2 = N\,.
\end{equation}
The GPE is an increasingly accurate approximation to the underlying quantum field theory, for instance, in the formal limit $N\rightarrow\infty$,  $g_{3}\rightarrow 0$ with $N g_{3}$ \colr{and the system volume} held constant.

We consider the external potential, $ V({\bf x})$, of the form
\begin{equation}
V({\bf x})={1\over 2} m (\omega_x ^2 x^2+\omega_{\perp} ^2 r_{\perp}^2)+V_0 \sin^2\left(\frac{\pi x}{d_L}\right).
\label{e2}
\end{equation}
Here $V_0$  and $d_L(=\lambda/2)$  are the depth and the periodicity of the optical lattice. If the harmonic confinement is much stronger in the transverse than in the longitudinal direction $(\omega_{\perp}\gg \omega_x)$ the GPE can be transformed into a one-dimensional form

\begin{eqnarray}
i\hbar {d\psi\over dt}
=\left(-{\hbar^2\over 2 m}{\frac{\partial^2}{\partial x^2}}+V({x})+g_1|\psi|^2 \right ) \psi,
\label{e1}
\end{eqnarray}
with an effective atom-atom interactions $g_1=2 a \hbar \omega_{\perp}$. 
Furthermore,
when the depth of the optical lattice is much larger than the chemical potential of the atoms, one can employ the tight-binding approximation. By expressing the condensate wave function $\psi(x)$ as a superposition of Wannier functions localized within each potential well of the lattice, one may obtain the tight-binding version of the GPE known as the discrete nonlinear Schr$\ddot{\text o}$dinger equation (DNLSE) \cite{SME02},
\begin{equation}
i{\partial\over\partial t}\psi_{\idx}=-J(\psi_{\idx+1}+\psi_{\idx-1})+(V_{\idx}+\chi \abs{\psi_{\idx}}^2)\psi_{\idx}.
\label{e3}
\end{equation}

We employ periodic boundary conditions. This means that for $L$ lattice sites numbered as $\idx=0,1,\ldots L-1$, the sites $\idx=L$ actually refers to the site $\idx=0$ and $\idx=-1$ to the site $\idx=L$. Physically this corresponds to a ring lattice. For other types of boundary conditions there obviously have to be some changes in the results. For instance,  in case of hard-wall boundary conditions a state of the atoms that would propagates around a ring will instead reflect back from the ends of the lattice. However, such variations of the theme will not be discussed further in this paper.

The parameters $J$ and $V_\idx$ characterize the tunneling rate of the atoms from site to site and the external trapping potential, respectively, whereas $\chi$ is proportional to the atom-atom interactions. Henceforth we ignore the non-lattice potential in the direction of the lattice, and correspondingly set $V_\idx\equiv0$. The value of the interaction parameter $\chi$ depends on the normalization of the wave function, i.e., of the complex numbers $\psi_\idx$,
\begin{equation}
{\cal N} = \sum_\idx |\psi_\idx|^2.
\end{equation}
Differently from both the GPE~\eq{ee1} and Ref.~\cite{Shrestha}, here we normalize the state to the one, ${\cal N}=1$. Even though the numbers $|\psi_l|^2$ then equal the fractions of the atoms residing at the sites $l$,  in what follows we nonetheless call them populations, or downright atoms numbers. The interaction coefficient is expressed in terms of the (actually three-dimensional) unit-normalized wave function of an atom in one (actually three-dimensional) potential well of the lattice $w({\bf x})$ as
\begin{equation}
g = \frac{4\pi\hbar a}{m} \int d^3x |\,w({\bf x})|^4,\quad \chi = \frac{N}{L}g,
\label{CHIDEF}
\end{equation}
which explicitly shows the difference between the inherently atomic coupling constant, $g$, and the coupling constant appearing in the DNLSE, $\chi$. We frequently regard the atomic coupling $g$ as variable, which might be achieved by employing a Feshbach resonance. The DNLSE becomes increasingly accurate, for instance, in the asymptotic limit $N\rightarrow\infty$ with $\chi$ and $L$ held constants.
Unless it is explicitly stated otherwise, we assume here that $J>0$ and the atom-atom interaction is repulsive, $\chi\geq 0$.

In the absence of the remnant trapping potential and nonlinearity Eq.~(\ref {e3}) may be solved with a plane-wave ansatz 
\beq
\psi_\idx^0(t)=\sqrt\frac{1}{L}\,e^{i[p\idx- \omega(p) t]},
\label{PLANEWAVE}
\eeq
 giving the dispersion relation 
 \beq
 \omega(p)=-2 J\cos p\,.
 \eeq
The boundary conditions quantize the lattice momentum $p$ to the values
\beq
p = \frac{2\pi}{L} P
\eeq
where $P$ is an integer that may be chosen to lie in the interval $[-\frac{L}{2},\frac{L}{2})$. For notational convenience we always take the number of lattice sites $L$ to be even.

When the interaction is switched on, the constant-amplitude plane waves are still stationary solutions to Eq.~(\ref{e3}) but with a modified dispersion relation 
\beq
\omega(p)=-2 J\cos p+{\frac{\chi}{L}. }
\label{DR}
\eeq
Besides these extended-wave solutions, the nonlinear system may also admit stationary solutions, of the form $\psi_\idx(t) = \psi_\idx(0)e^{-i\mu t}$, which are localized in space \cite{Dauxois}. These solutions, so-called gap solitons,  usually have an energy that lies outside of the linear band spectrum. In the continuum model solitonic solutions of the nonlinear Schr\"odinger equation can be found in closed form by using the inverse scattering method \cite{Akhmediev}. However, the discrete system has fewer constants of the motion and is not integrable as such, so that one has to rely on numerical techniques for exact solutions.

\subsection{Modulational instability}

To study the stability of a the plane wave solution~\eq{PLANEWAVE} and \eq{DR}, we introduce a perturbation around the steady state \cite {Wu,SME02},
\begin{equation}
\psi_\idx(t)=\psi_\idx^0(t)[1+u_q e^{i(q \idx-\Omega_q t)}+v_q^*e^{-i(q \idx-\Omega_q^* t)}],
\label{e12}
\end{equation}
where $q$ and $\Omega_q$ are the quasimomentum and the frequency of the small excitation relative to the initial unperturbed steady state, and $u_q$, $v_q$ are assumedly small mode amplitudes. By the periodic boundary conditions the possible excitation modes $q$ are also quantized and may be characterized by an integer $Q\in[-L/2,L/2)$, so that
\beq
q = \frac{2\pi Q}{L}\,.
\eeq
$Q=0$, though, corresponds to a change in the normalization of the unperturbed state $\psi_\idx^0$, and does not qualify as an excitation.

 After inserting Eq.~(\ref {e12}) in Eq.~(\ref {e3}) with $V_\idx=0$, and expanding to the lowest nontrivial order in $u_q$ and $v_q$ we get the following matrix equation,
\begin{equation}
i{d{U_q}\over dt}={\cal M}{U_q},
\label{e13}
\end{equation}
where $U_q$ is the vector $[u_q,v_q]^T$ and ${\cal M}$ is a $2\times 2$ matrix with the elements
\begin{eqnarray}
&&{\cal M}_{11}=\frac{\chi}{L} +4 J \sin \frac{q}{2}~\sin(\frac{q}{2}+p),\nonumber\\
&&{\cal M}_{22}=-\frac{\chi}{L} -4 J\sin \frac{q}{2}~\sin(\frac{q}{2}-p),\nonumber\\
&&{\cal M}_{12}=-{\cal M}_{21}^q=\frac{\chi}{L}.	
\label{e14}
\end{eqnarray}
The eigenvalues of $\cal M$ give the small-excitation frequencies,
\begin{eqnarray}
\Omega_q&=& 2J \sin p \sin q \pm \Gamma_q;\label{e15}\\
\Gamma_q&=&\sqrt{J \cos p \sin ^2\frac{q}{2} \left[2 J \cos (p) \sin
   ^2\frac{q}{2}+\frac{\chi }{L}\right]}\,,
\label{GAMMADEF}
\end{eqnarray}
whereas the eigenvectors give the corresponding mode amplitudes,
\begin{equation}
u_q=\sqrt{\frac{4 J \cos p\, \sin ^2\!{\displaystyle\frac{q}{2}}\pm\Gamma_q +\displaystyle{\frac{\chi}{L}}
   }{2\Gamma_q }}\,,
\label{ee18}
\end{equation}
\begin{equation}
v_q=\sqrt{\frac{4 J \cos p\, \sin ^2\!{\displaystyle\frac{q}{2}}\mp\Gamma_q +\displaystyle{\frac{\chi}{L}}
   }{2\Gamma_q }}\,.
\label{ee19}
\end{equation}

As long as the quantity $\Gamma_q$, Eq.~\eq{GAMMADEF} is real, the frequency of the excitation mode $\Omega_q$ is also real and the amplitudes $u_q$ and $v_q$ are chosen to satisfy the normalization conditions $|u_q|^2-|v_q|^2=\pm 1$, the sign depending on the mode. The reason for this normalization is that in quantum theory a unit-normalized state
\beq
 {\psi_\idx(p,q) = \frac{1}{\sqrt L}(u_q e^{i(p+q)\idx} + v^*_q e^{i(p-q)\idx})}
\eeq
then represents one bosonic elementary excitation of the condensate. There are two excitation modes corresponding to the $\pm$ signs in Eqs.~\eq{e15}-\eq{ee19} for each quasimomentum $q$, but only those with positive normalization give physical small-excitation modes. The modes corresponding to negative normalization are artifacts of the present Bogoliubov type analysis, and are henceforth ignored. For more details on the linear stability analysis and its relation to quantum field theory consult, e.g., Refs.~\cite{JAV98,JAV01}.

Since we are dealing with the repulsive atom-atom interactions, $\chi\geq 0$, all small-excitation frequencies $\Omega_q$ are real for $|p|\le \pi/2$. However, the existence of complex eigenvalues cannot be ruled out in the interval $|p|\in (\frac{\pi}{2},\pi]$, depending on the values of $J,~\chi,~q$ and $p$. If for a given $p$  there exists a mode $q$ such that the quantity $\Gamma_q$ is imaginary, there is a small excitation that grows exponentially. This signals a dynamical instability of the steady-flow mode $p$. For a dynamically unstable excitation mode $q$ we have  $|u_q|^2-|v_q|^2\equiv0$, so that unstable modes cannot be normalized to one~\cite{Wu}.

\begin{figure}
\includegraphics[clip,width=8cm]{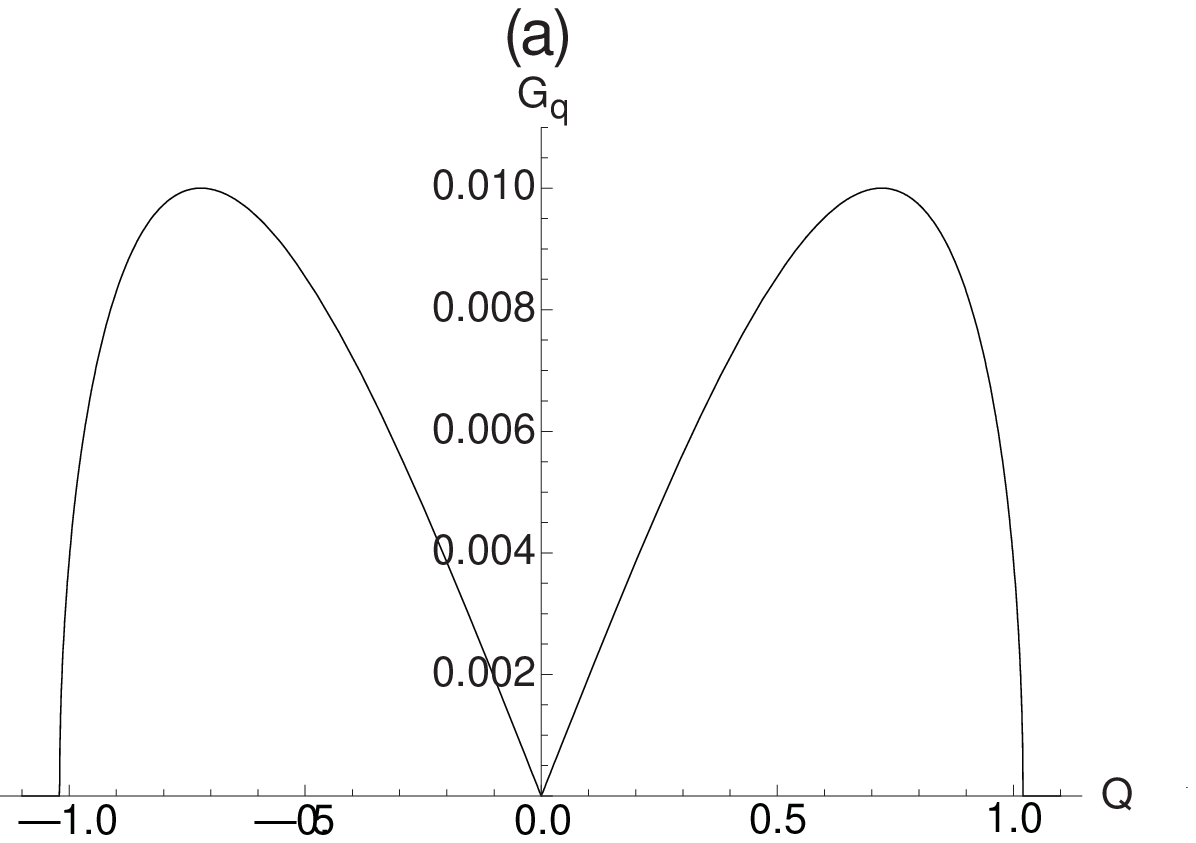}
\includegraphics[clip,width=8cm]{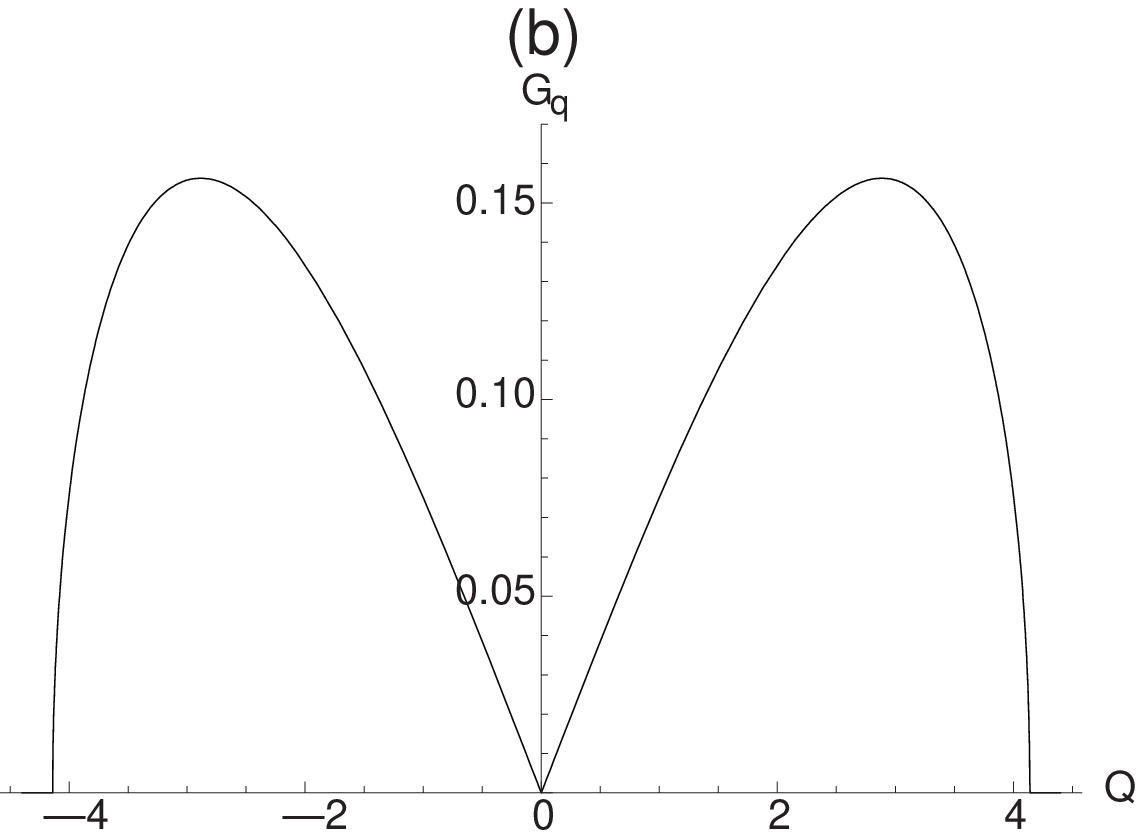}
\caption{Coefficient of exponential gain ${G_q}$ for an excitation mode as a function of the parameter $Q$ related to quasimomentum $q$ by $Q=Lq/(2\pi)$. The parameters characterizing the interaction strength are $\Lambda=0.32$ (a) and $\Lambda=5.0$ (b). Modulational instability is only possible for $G_q>0$, and excitation modes may only occur at nonzero integer values $Q$.}
\label{f2}
\end{figure}

Figure~\ref{f2} shows the gain coefficient, 
\beq
G_q=\frac{|\Im\Omega_q|}{2J},
\eeq
for two \colr{cases (a) $\Lambda=0.32$, and (b) $\Lambda=5.0$, where the dimensionless interaction strength $\Lambda$ is defined to be
\begin{equation}
\Lambda=\frac{\chi}{2J}\,.
\label {ep1}
\end{equation}
Here we consider} a lattice with $L=32$ sites for $p=\pi$. Figure~\ref{f2}(a) reveals a single pair of unstable mode with $Q=\pm1$, whereas Fig.~\ref{f2}(b) shows four pairs of unstable modes.

Analytically, for any $|p|$ greater than ${\pi/ 2}$ and in the limit of large $L$, the eigenfrequency for the excitation mode $Q$ will be complex if
\begin{equation}
\Lambda>|\cos(p)|~\frac{\pi^2 Q^2}{L}\colr{\,.}
\label{e16}
\end{equation}
The corresponding expression for the gain coefficient is
\beq
G_q = \frac{2\pi Q \sqrt{|\cos p|(L \Lambda - \pi^2Q^2 |\cos p|)}}{L^2}\,,
\eeq
and the characteristic time scale for the dynamical instability $\tau_q$ is given by
\beq
2J\tau_q = \frac{1}{G_q}\,.
\label{TIMESCALE}
\eeq

The flow with the quasimomentum $p$ is dynamically unstable if the inequality~\eq{e16} is true at least for the longest-wavelength excitation with $Q=1$. The critical interaction strength approaches zero when the number of lattice sites $L$ tends to infinity, whether for a fixed atom number or if the atom density $\propto N/L$ is held constant. In this way any flow with $|p|>\pi/2$  will turn unstable with $L\rightarrow \infty$. The imaginary part of the complex eigenfrequency as well as the corresponding eigenvectors are the same for the modes $Q$ and $-Q$ [see Eqs.~(\ref{ee18}), (\ref{ee19})], so that we regard these two modes as equivalent as it comes to the instability.

\section{\label{TE}Time Evolution and Pulsating Instability}

We carry out numerical simulations on the DNLSE to study the growth of the unstable excitation mode in a lattice for a suitable range of interaction parameters. For a given number of lattice sites $L$ and flow momentum $p$, the number of unstable modes in the linear stability analysis depends only on the interaction parameter $\Lambda$, Eq.~\eq{ep1}. Here we focus on long-wavelength excitations in the limit of weak atom-atom interactions. Two numerical methods have been used for the time evolution, an unconditionally stable Crank-Nicholson type algorithm \cite{NR} and a sixth-order accurate FFT split operator algorithm that works in the same way for DNLSE as is discussed in Ref.~\cite{Javanainen} with the ordinary GPE.

\subsection{Single unstable mode}

\begin{figure}
\includegraphics[clip,width=8 cm]{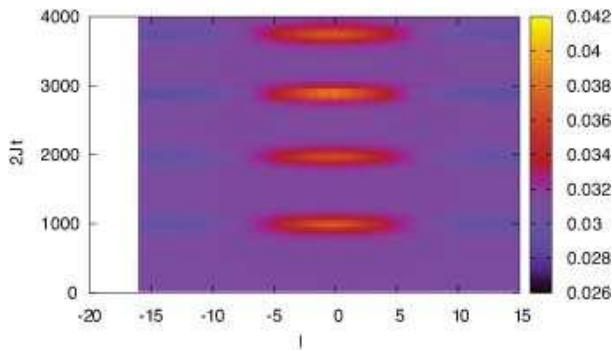}
\caption{(Color online) Collapse and revival of the density pulse in the BEC evolution when the lattice starts from a dynamically unstable state in a density plot for the populations of the sites $|\psi_l|^2$, the lighter shades corresponding to larger populations. The parameters are $\Lambda=0.48$, $p=\pi$, and $L=32$. These parameters correspond to a single unstable mode, as per linear stability analysis. The instability drives the initial homogeneous atom distributions into a density peak that subsequently disperses back to the initial state, and the process repeats.}
\label{f3}
\end{figure}

A perusal of the condition (\ref{e16}) shows that the range of interaction strengths where the $Q=1$ mode is unstable but the $Q=2$ mode is not is given by

\begin{equation}
 |\cos p|~\frac{\pi^2}{L} < \Lambda \le 4 |\cos p|~\frac{\pi^2}{L}.
\label{e17}
\end{equation}
Figure~\ref{f3} shows a typical density plot of the time evolution of the BEC initially prepared in the plane wave state at the edge of the Brillouin zone ($p=\pi$) seeded with random Gaussian noise, for the number of lattice sites $L=32$ and the interaction strength $\Lambda=0.48$. This value of $\Lambda$ corresponds to one unstable mode in the linear stability analysis. Although a tiniest amount of noise (either in real experiments or in numerical simulations) is sufficient to trigger the instability, an external noise of amplitude $\sim10^{-4}$ is added to speed it up. The same applies to all of our demonstrations of the pulsating instability in multisite lattices up to Sec.~\ref{TWA}, where we adopt a model for the quantum noise instead. It has been tested in a number of runs that the time for the  onset of the instability for fixed values of the other parameters depends logarithmically on the amplitude of the added noise, as it should.

\begin{figure}
\includegraphics[clip,width=1.0\linewidth]{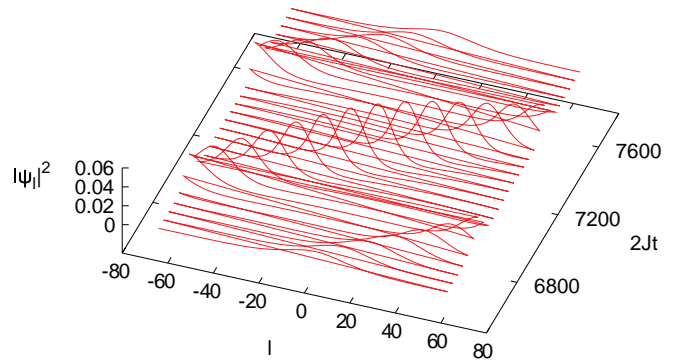}
\caption{(Color online) Density evolution of the BEC for the initial flow momentum different from $\pi$. Here the parameters are $L=128,~\Lambda=0.25$ and $p=15\pi/16$. }
\label{f4}
\end{figure}

Figure~\ref{f4} depicts snapshots of a pulse that moves during its formation from the initial flow state with quasimomentum $p=15~\pi/16$. In this figure we take a larger lattice with $L=128$ sites and the interaction strength is $\Lambda=0.25$, so that there is still one and only one unstable mode.  The pulsating behavior of the peak still persists but the peak moves. By virtue of the periodic boundary conditions a pulse that goes over the right edge will reappear at the left edge of the lattice.

\begin{figure}
\includegraphics[clip,width=1.0\linewidth]{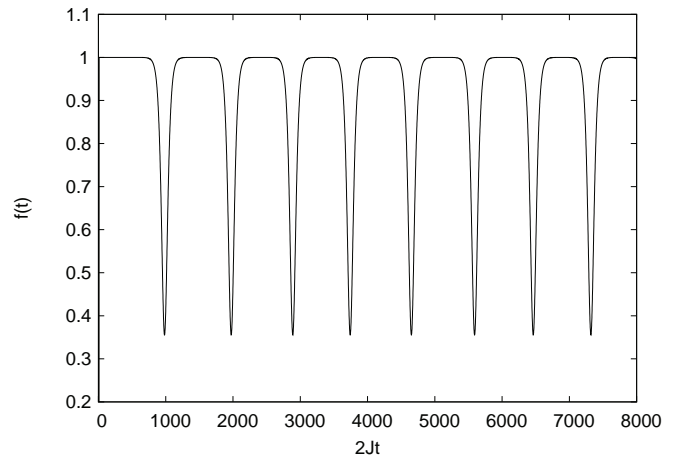}
\caption{Fraction of the initial state $f(t)$ in the state of the lattice plotted as a function of time. Here the parameters are $L=32, ~\Lambda=0.48$ and $p=\pi $.}
\label{f5}
\end{figure}

The pulsating behavior also manifests in the fraction of the initial state $\psi_\idx(0)$ remaining in the state of the lattice  at a given time,
\begin{equation}
f(t)=\left|\sum_\idx \psi_\idx ^*(0) \psi_\idx(t)\right|^2.
\end{equation}
In Fig.~\ref{f5} we plot the overlap, $f(t)$, as a function of time for the same parameters as in Fig.~\ref{f3}. It is revealed that the instability drives the system far from, and subsequently brings it back to, the original unstable steady state, and the process repeats. Each dip in the plot represents formation of a pulse during the course of time. It is also noted that the quantity $f(t)$ does not vanish all the way to zero, indicating that the pulsed state is not orthogonal to the initial steady state. Furthermore, a closer inspection of this plot shows that the subsequent peaking events are not strictly periodic; the interval between the dips varies slightly, implying a quasi-periodic phenomenon.

By analyzing data of this kind a number of observations emerge. First, in contradiction to the common belief that the instability may develop into an irregular dynamics, here it causes the atoms to pile up in a single-peaked distribution of the site populations $|\psi_\idx|^2$. Furthermore, upon continued time evolution, the system returns very close to the initial unstable state, again pulsates to a peak, and so on. We have nearly periodic peaking and recurrences to the unstable initial state. Second, for the initial flow state $p\neq \pi$, the pulse also moves with a velocity that turns out to be close to the group velocity of the carrier wave, $v_g =2J\sin p$.
Third, the peak may occur at any lattice site. It is the random noise that seeds the position of the peak. This is in accordance with translational invariance of the lattice: A lattice-translated pulsed solution is also a solution of the DNLSE, and there is no preferred lattice site for the occurrence of the pulse. Fourth, for weak initial noise the dependence of the pulsation phenomena on the initial noise is otherwise weak. For instance, even if the timing of the first pulse depends logarithmically on the noise amplitude and the spacing of the subsequent peaks is also affected, the dominant time scale is evidently the time scale of the instability from Eq.~\eq{TIMESCALE}. In Fig.~\ref{f5} the peak spacing equals approximately $10\,\tau_q$.

\subsection{Multiple unstable modes}

For a lattice with a large number of sites the one-peak condition~\eq{e17}  is highly impractical because the interaction strength $\Lambda$ needs to be extremely small and the pulse revival period is correspondingly long. For instance, the shortest possible time scale for the dynamical instability in the one-peak case, $\tau$, is determined from Eqs.~\eq{e17} and~\eq{TIMESCALE} as $2J\tau = L^2/(2\sqrt{3}\pi^2|\cos p|)\propto L^2$. 

\begin{figure}
\includegraphics[clip,width=1.0\linewidth]{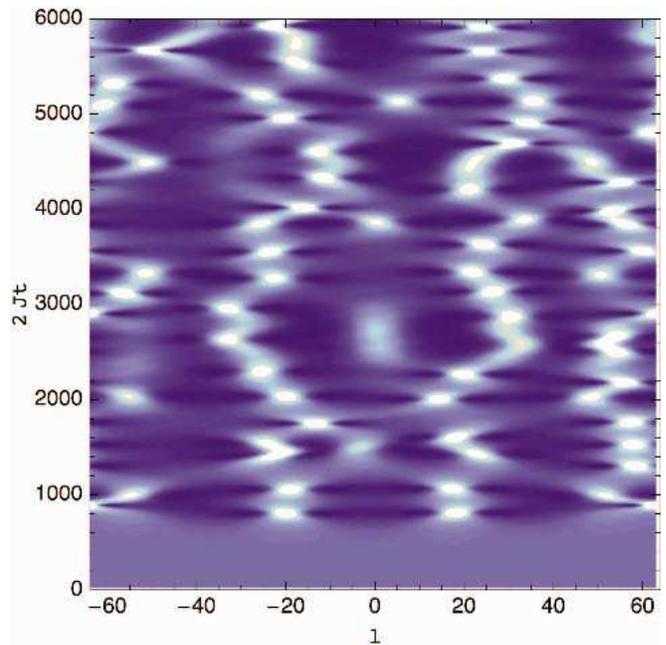}
\caption{(Color online) Density plot of the populations $|\psi_\idx|^2$ in the case where there are four unstable modes. The parameters are $L=128$, $\Lambda=2.0$, and $p=\pi$. Lighter shading represents higher site populations. }
\label{f6}
\end{figure}

On the other hand, the condition that exactly $Q$ modes are unstable is obtained from Eq.~(\ref{e17}) in the form
\begin{equation}
 Q^2|\cos p|~\frac{\pi^2}{L} < \Lambda \le (Q+1)^2|\cos p|~\frac{\pi^2}{L}.
\label{e18}
\end{equation}
For reasonable interaction strengths there might be more than one unstable mode. Figure~\ref{f6} is a typical representative of the dynamics of the BEC with multiple unstable modes. Here we take $N=128$, $\Lambda=2.0$, and $p=\pi$. Each bright spot represents a pulse. As before, the right edge of the plot wraps around to the left edge by virtue of the periodic boundary conditions. Equation~\eq{e18} gives five unstable modes, while most of the time we see three or four pulses in this example. However, these pulses are not independent. Presumably because of nonlinear mode-mode interaction, they move around, join and split as they collapse and revive.

.

\subsection{Evolution in Fourier space}

The recurrences observed in the peaking events in the DNLSE resemble the energy recurrences in the Fermi-Pasta-Ulam (FPU) problem \cite{Fermi}. The FPU model deals with the evolution of a lattice chain with nonlinear interactions between the nearest-neighbor atoms when  initially a single low-energy mode is excited. For a time scale much longer than the time period of the normal modes, the energy is well localized to the given excited mode, while the amplitudes of the higher-energy modes decay exponentially as a function of the energy difference from the initially excited mode. For a longer time scale it has also been noticed that recurrence of the initial excited mode is possible.

The pulsating behavior of the density distribution of the BEC atoms in the lattice can be viewed as a similar recurrence phenomenon as observed in the FPU model. We have started with a steady state for a given flow quasimomentum ($p>\pi/2$) and a suitable nonlinear interaction strength to  trigger the instability in the system. Ergodicity immediately suggests that the energy initially fed into a single mode should distribute evenly between all Fourier modes. However, the excitation amplitudes of the modes other than the mode corresponding to the initial steady state seems to decay exponentially with the index $Q$. The energy localization to a few Fourier modes in a nonlinear system has been studied also in other systems~\cite{Dauxois}. The existence of discrete breathers in a nonlinear lattice system is an example. Recently, energy localization in Fourier space in a so called `q-breather' has been investigated in \cite{Flach}.

\begin{figure}
\includegraphics[clip,width=1.0\linewidth]{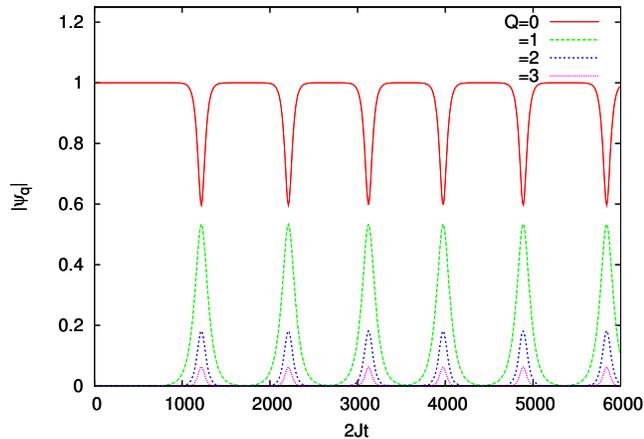}
\caption{(Color online) Time evolution of the Fourier modes $\psi_q$ of the lattice. The parameters are $L=32$, $p=\pi$ and $\Lambda=0.48$. The mode $Q=0$ corresponds to the initial steady state while modes $Q=1,2,3\dots$ are the low-lying excitations. Only the modes with $Q>0$ are shown.}
\label {f7}
\end{figure}

Figure \ref{f7} shows the time evolution of the Fourier modes $\psi_q$ of DNLSE for the parameters $ \Lambda=0.48,~N=32$. Only a few components are seen to be excited, as the amplitudes of the higher-energy modes are suppressed exponentially. The dominant Fourier components are $P$ and $P\pm1$ which implies that the excitation modes that go unstable should have the indices $Q=\pm 1$ with respect to the initial steady state, as expected.

\section{\label{DW}Double Well Analogy}
In order to explain qualitatively the pulsating behavior of the density distribution of the BEC in the lattice, we study a coupled double-well system. Writing $\psi_{0,1}=|\psi_{0,1}|e^{i\phi_{0,1}}$, the dynamics of such a system can be described by a pair of equations \cite{Raghavan}

\begin{eqnarray}
&&\dot z(t)=-\sqrt{1-z^2(t)}\sin\phi(t), \nonumber\\
&&\dot \phi(t)=\Lambda z(t)+ {z(t)\over \sqrt{1-z^2(t)}} \cos\phi(t),
\label{IV1}
\end{eqnarray}
where $z=|\psi_{1}|^2-|\psi_{0}|^2$ and $\phi=\phi_{1}-\phi_{0}$ are the fractional population imbalance and the relative phase between the two wells. The normalization again is $|\psi_{0}|^2+|\psi_{1}|^2=1$. The correct evolution for $z$ and  $\phi$ is obtained from the Hamiltonian
\begin{equation}
H={\Lambda z^2\over 2}- \sqrt{1-z^2}~\cos\phi
\label{IV2}
\end{equation}
by regarding these quantities as canonical conjugates. The Hamiltonian (energy) is a constant of the motion and the double-well system is therefore integrable. The evolution of the system keeps it on a constant-energy contour in the space with $z$ and $\phi$ as the axes.

By inspection it can be checked that the fixed points of  Eq.~(\ref{IV1}) are $ z=0 , \phi=n \pi$, where $n$ is an integer. A potentially unstable steady state in the multiwell system corresponds in  the two-well system to the steady state $z=0$, $ \phi=\pi$. The behavior of the orbits near this equilibrium point can be examined by using linear stability analysis as before. It can be easily verified that the state $(0,\pi)$ is stable for the values $\Lambda\leq 1$, and unstable otherwise.

\begin{figure}
\includegraphics[clip,width=1.0\linewidth]{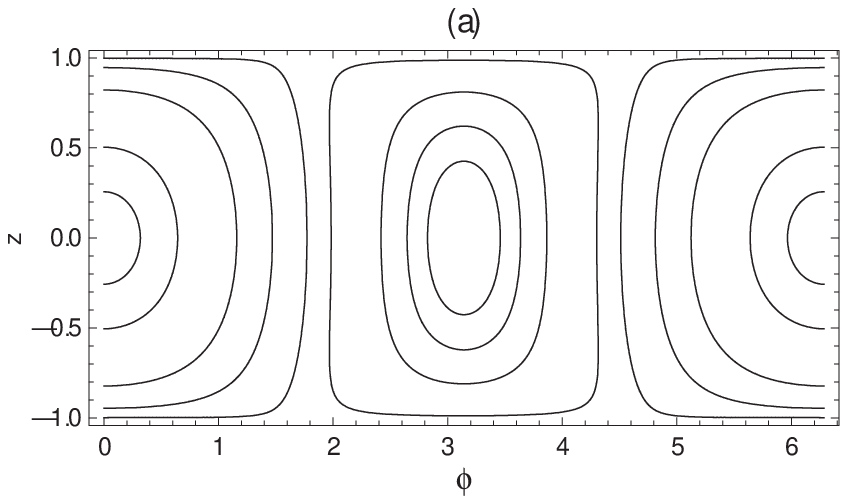}
\includegraphics[clip,width=1.0\linewidth]{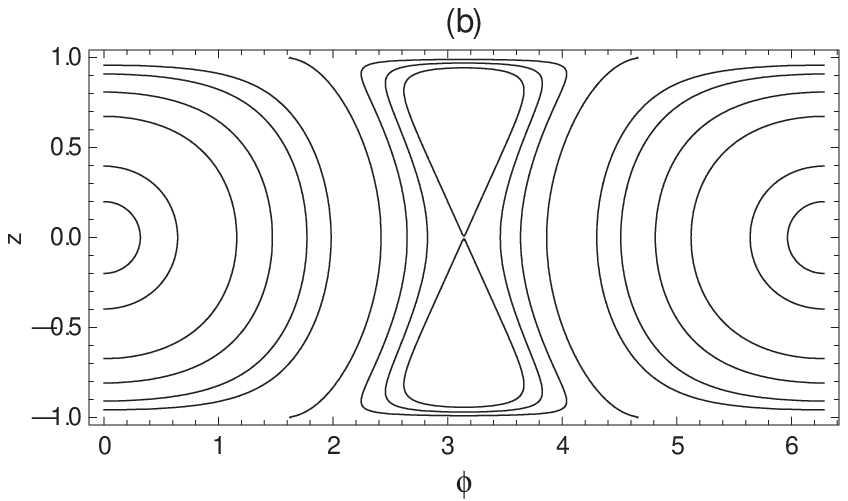}
\caption{Representative constant-energy contours of the two-site Hamiltonian~\protect\eq{IV2} in the $(\phi,z)$ plane for the interaction strengths $\Lambda=0.5$~(a) and $1.5$~(b). }
\label{f8}
\end{figure}

In Fig.~\ref{f8} we show constant-energy contours of the two-well system for $\Lambda=0.5$ (a) and $1.5$ (b). We have drawn the $\phi$ axis from $0$ to $2\pi$ so that the potentially unstable fixed point $(0,\pi)$ lies at the centers of the panels.  For $ \Lambda=0.5$ the potentially unstable steady state is an elliptic fixed point and the time evolution takes the system periodically around this point. At $\Lambda=1$ the elliptic fixed point bifurcates, and for $\Lambda=1.5$ there is a homoclinic orbit with the emergence of two symmetric off-centered elliptic fixed points.  Thus, starting in the vicinity of what used to be the potentially unstable steady state, the system takes off in an unstable direction along the homoclinic orbit and goes around one of the bifurcated elliptic fixed points.

\begin{figure}
\includegraphics[clip,width=1.0\linewidth]{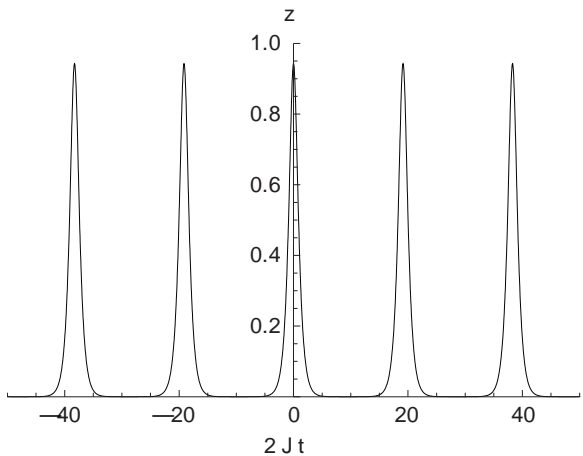}
\caption{Time evolution of the population imbalance $z$ given by an exact analytical solution of  Eqs.~(\ref{IV1}) for the parameter $\Lambda=1.5$ and the value of the Hamiltonian $H=1.0$.}
\label{f9}
\end{figure}

The double-well system also admits an analytic solution in terms of Jacobian elliptic functions. In Fig.~\ref{f9}  we plot such a solution for $z(t)$ for the parameter $\Lambda=1.5$ and the conserved value of the Hamiltonian $H=1.0$.  The double-well system is then unstable in linear stability analysis and the energy corresponds to an energy contour close to the homoclinic orbit, see Fig.~\ref{f7}(b). The oscillations in the population imbalance bear a striking resemblance to Fig.~\ref{f5} with the plot of the overlap in the multiwell lattice, and may be viewed as an analogue of the pulsating instability.

We have presented  two different views on the pulsating instability in the two-site system, energy contours and time evolution of the population imbalance, but they describe the same physics. The dynamically unstable system performs periodic oscillation where the system repeatedly recedes far away from the unstable state and subsequently returns.

The multisite system basically shares the behavior of the two-site system in a multi-dimensional phase space: Starting from random noise in the neighborhood of an unstable steady state, it evolves away from, and returns to, the initial state, and the process then repeats. These periodic recurrences occur in a $2L$-dimensional phase space on a constant-energy surface in full analogy with the two-site system. The two-site system is strictly periodic since there is no motion out of the constant-energy surface, a one-dimensional curve. However, in the multiwell case the dimension of the constant-energy surface is $2L-1$. Our pulsating instability strongly suggests that the multiwell system stays close to the homoclinic orbit while it evolves. However, depending on the initial noise it still has a large state space to explore, and the motion may deviate slightly from the homoclinic orbit. Upon looping around one of the stable fixed points, the multisite system therefore does not have to return to exactly where it started from. This might account for the slight variations in the period of the pulsations.

In short, we attribute the pulsating dynamics in the multiwell system to a remnant of integrability. The multiwell system evolves along a similar homoclinic orbit that exists in a two-well system, and the motion away from the homoclinic orbit remains bounded. 
\commentout{\colr{As the double-well dynamics of a BEC can generally be strongly influenced by quantum dissipation \cite{RUO98b,ZAP03}, we will investigate in next Section the effect of quantum fluctuations on the pulsating phenomenon in an optical lattice.}}

\section{\label{TWA}Truncated Wigner approximation}

\commentout{
In section II we discussed the dynamics of a BEC within the classical mean-field theory. The GPE can constitute a very accurate modeling of a weakly interacting BEC. In optical lattices, however, the kinetic energy is represented by the hopping of the atoms between adjacent lattice sites, and the hopping amplitude can be significantly reduced in deep lattices. Quantum correlations and fluctuations are a matter of balance between atom-atom interactions and hopping, and may be correspondingly enhanced. In the following we include quantum fluctuations in the dynamics in a lattice using a stochastic phase space method, the truncated Wigner approximation (TWA).
}

The GPE can constitute a very accurate modeling of a weakly interacting BEC. In Secs.~\ref{TM}--\ref{DW}
we discussed the dynamics of a BEC within the classical mean-field theory, including
double-well and multi-well systems. We know in the case of double-well systems,
however, that when the tunneling barrier becomes large, quantum and thermal
fluctuations can considerably influence the atom dynamics \cite{RUO98b}.
Similarly, in optical lattices the kinetic energy is represented by the hopping of the
atoms between adjacent lattice sites, and the hopping amplitude can be significantly
reduced in deep lattices. Quantum correlations and fluctuations are a matter of
balance between atom-atom interactions and hopping, and may be correspondingly
enhanced. In the following we include quantum fluctuations in the dynamics
in a lattice using a stochastic phase space method, the truncated Wigner
approximation (TWA).

TWA was introduced for multi-mode dynamics  in nonlinear optics~\cite{DRU93}. Details how to implement TWA in different atomic BEC systems may be found, e.g., in Refs.~\cite{Steel,SIN02,GAR02,ISE06}, studies of zero- and finite-temperature nonequilibrium dynamics in 1D lattice systems are documented in Refs~\cite{POL03c,ISE05,ISE06,RUO05,Ferris,TUC05}, and  the effects of dynamical instabilities in lattices have been explicitly addressed in Refs.~\cite{POL03c,RUO05,Ferris}.
In the TWA one neglects the third-order derivatives in the generalized Fokker-Planck like equation for a Wigner distribution function~\cite{Gardiner}. This permits a nonlinear stochastic differential equation for a classical field that represents the quantum field. Here the equation is just the DNLSE.

In our present case of TWA quantum fluctuations and correlations enter only through the \colr{stochastic} initial state of the classical field. Our emphasis on quantum fluctuations is different from typical finite-temperature dominated TWA approaches in higher dimensions~\cite{BIS08}. In an attempt to capture quantum mechanics as accurately as practicable, we therefore pick the initial states of the wave function using the Bogoliubov approximation.

In our numerical simulations we consider a stable stationary superfluid flow that is instantaneously, at $t=0$, rendered dynamically unstable by changing atom-atom interactions. We investigate the effect of quantum fluctuations by considering
two different types of initial states: a noninteracting BEC with all atoms in the same one-particle state, and an
interacting system in which the atom-atom interactions force  fraction of the atoms out of the condensate. We first {generate} a \colr{stochastic} initial state  $\psi_\idx^W(t=0)$  accordingly, then the TWA dynamics follows from the DNLSE
\begin{equation}
i{\partial\over\partial t}\psi_{\idx}^W=-J(\psi_{\idx+1}^W+\psi_{\idx-1}^W)+\chi \abs{\psi_{\idx}^W}^2\psi_{\idx}^W\,.
\label{V1}
\end{equation}
This process is repeated for a number of initial states. Two types of results are of interest: Individual trajectories $\psi_\idx^W(t)$ represent the outcomes of individual experiments, and appropriate averages over the collection of trajectories are used to calculate pertinent quantum expectation values. The present TWA formalism is very similar to the one used in Refs.~\cite{ISE05,ISE06,RUO05}, except that in each TWA realization we fix the total atom number \cite{MAR08}.

\subsection{Initial state}

In order to describe the generation of the stochastic initial states  for the TWA dynamics we begin with the quantum version of the Bogoliubov approximation. We again consider the stationary solution for a moving plane wave as in Eqs.~\eq{PLANEWAVE} and~\eq{DR}. The fluctuations around the stationary solution are governed by the decomposition of the atom field operator
\begin{equation}
\hat{\psi}_{\idx}(t)=\psi_{\idx}^0(t) \hat\alpha_0 +\delta \hat{\psi}_\idx(t)\,,
\label{V11}
\end{equation}
where the total number of condensate atoms,
\begin{eqnarray*}
N_c=\langle\hat\alpha_0^\dagger \hat\alpha_0\rangle,
\end{eqnarray*}
is assumed much larger than one and $\delta \hat{\psi}_\idx(t)$ is supposedly \colr{``small.''} Heretofore we make a difference between total number of atoms $N$, number of condensate atoms $N_c$, and number of noncondensate atoms $N_n$. Analogously to our earlier classical Bogoliubov treatment, the fluctuation part of the atom field operator, $\delta\hat{\psi}_{\idx}(t)$, can be written in terms of quasiparticle  operators $\hat{\alpha}_q$ and $\hat{\alpha_q}^\dagger$ as
\begin{equation}
\delta\hat{\psi}_{\idx}(t)\!=\!\frac{1}{\sqrt L}\sum_{q\neq0} (u_q\hat{\alpha}_q \text e^{[i(p+q )\idx-i\Omega_q t]}+v_q^*\hat{\alpha}_q^\dagger\text e^{i[(p-q)\idx +i\Omega_q^* t]})\,.
\label{V12}
\end{equation}
Within the Bogoliubov approximation the quasiparticles make a bosonic ideal gas, or alternatively,  the operators $\hat{\alpha}_q$ may be viewed as the lowering operators for a collection of independent harmonic oscillators. The normal mode frequency $\Omega_q$ and the quasiparticle amplitudes $u_q$, and $v_q$ are given by Eqs.~(\ref{e15}), (\ref{ee18}), and~(\ref{ee19}); only the excitation modes with positive normalization are used. The unit normalizations for both the unperturbed plane wave $\psi_\idx^0$ and the amplitudes of the excitation modes $u_q$, $v_q$ were fixed earlier so that these quantities work as given also in the present quantum version of Bogoliubov theory.

The expectation value of the number of non-condensate atoms in the Bogoliubov theory is
\begin{equation}
\bar N_{n}=\sum_{q}(|u_q|^2+|v_q|^2) \langle\hat\alpha_q^\dagger \hat\alpha_q \rangle+\sum_{q}|v_q|^2,
\label{V13}
\end{equation}
with
\begin{equation}
\langle \hat\alpha_q^\dagger \hat \alpha_q \rangle =[e^{\Omega_q/{k_B T}}-1]^{-1}.
\label{V14}
\end{equation}
At $T=0$ we have $\langle \hat\alpha_q^\dagger \hat \alpha_q \rangle=0$, and the non-condensate atom number is simply
\begin{equation}
\bar N_{n}=\sum_{q}|v_q|^2.
\label{V15}
\end{equation}

In order to construct the initial state of the lattice for TWA evolution we replace the quantum operators $(\hat{\alpha}_q,\hat{\alpha}_{q'}^\dagger) $ in Eq.~(\ref{V12}) by complex stochastic variables $(\alpha_q,\alpha_{q'}^*)$ obtained by sampling the corresponding Wigner distribution function. Our formalism
follows Ref.~\cite{ISE06}, except that here we fix the total atom number \colr{\cite{MAR08}}. The Wigner function at $T=0$ is \cite{Gardiner}
\begin{equation}
W(\alpha_q,\alpha_q^*)=\frac{2}{\pi}\exp[-2|\alpha_q|^2]\,.
\label{V16}
\end{equation}
	The function $W(\alpha_q,\alpha_q^*)$ is a Gaussian and we easily find the average  number $\langle \alpha_q^* \alpha_q\rangle_W=\frac{1}{2}$ of excitations in each mode $q$ owing to quantum noise. Each unoccupied excitation mode exhibits uncorrelated Gaussian noise, distributed over the plane wave basis, and normalized to an average of a half an excitation quantum per mode. The attendant atoms provide seeding for scattering events in the TWA dynamics. However, Wigner functions correspond to symmetrically ordered expectation values of functions of the operators $\hat\alpha_q$ and $\hat\alpha^\dagger_q$, and in the end the expectations values  $\langle \alpha_q^* \alpha_q\rangle_W=\frac{1}{2}$  need to be subtracted to obtain normally-ordered quantum averages. For each stochastic realization, the number of non-condensate atoms therefore reads
\begin{equation}
N_{n}=\sum_{q}(|u_q|^2+|v_q|^2) (\alpha_q^* \alpha_q-\frac{1}{2})+\sum_{q}|v_q|^2.
\label{V17}
\end{equation}
This may fluctuate about the mean value $\bar N_n=\sum_{q}|v_q|^2$ for each realization, but correctly averages to $\bar N_{n}$.

Since the total atom number $N$ is fixed, the number of condensate atoms in each individual run is given by
\begin{equation}
N_{c}=N-N_{n}\,,
\label{V18}
\end{equation}
which also fluctuates around the mean value $\bar N_c=N-\colr{\bar N_n}$. In the initial state for each realization we then set
\begin{equation}
\alpha_{0}=\sqrt{ N_c +1/2}\,,
\label{V19}
\end{equation}
so that, in the place of Eq.~(\ref{V11}), we have the initial state for the TWA method
\begin{equation}
{\psi}_{\idx}^W (0)=\psi_{\idx}^0(0) \alpha_0 +\frac{1}{\sqrt{L}}\sum_{q\neq0} (u_q {\alpha_q} e^{i(p+ q)\idx }+v_q^*{\alpha_q}^* e^{i (p-q)\idx})\,,
\end{equation}
where $\alpha_q$ and $\alpha_q^*$ are \colr{stochastic variables in} each realization of the TWA. Note that, even though we consider a uniform system with a plane wave phonon basis and uncorrelated noise in the initial phonon modes, the fixing of the total atom number introduces long wavelength correlations in the system between the condensate mode and the excited quasiparticle modes~\cite{MAR08}.

\commentout{
\colr{
The results quoted thus far also illuminate situations where the classical DNLSE alone, as in Eq.~\eq{e3}, might
suffice for the description of the condensate evolution. Overall, the DNLSE can only provide an accurate model of a multi-mode dynamics if each relevant mode has an occupation number much larger than one. In the present case, we study as an initial state linearized fluctuations around a stationary superfluid flow over a large number of closely spaced excited modes, none of which exhibits a macroscopic occupation, even if the number of atoms is large, $N\gg1$. Consequently, we require that none of the weakly populated modes in the initial state play an important role, indicating that the number of atoms that the interactions force out of the condensate must be a small fraction of the total number of atoms, $\bar N_n\ll N$. 
Now, the expression~\eq{ee19} for the amplitudes $v_q$ shows that the interaction strength enters through the dimensionless parameter $\chi/J$. By virtue of our normalizations, the interaction parameter $\chi$, Eq.~\eq{CHIDEF}, is proportional to the number of atoms. If the number of atoms is increased while adjusting the strength of the atom-atom interactions in such a way that $\chi/J$ is held constant, the number of noncondensate atoms remains constant and the fractional error of the DNLSE decreases. 
Moreover, each mode in the TWA initial state is populated with an average of half a particle due to quantum noise. If the number of relevant modes participating in the dynamics is $\mathfrak{p}$, we then also require that in the classical limit $N \gg \mathfrak{p}/2$. While this number generally depends on the mode spacing and the dimensionality of the problem, it may be numerically tested, and the classical DNLSE limit with a sufficiently large $N$, for a given $\chi$, can be found.}}

Based on the Bogoliubov results it is interesting to estimate when the classical description
of the condensate dynamics from the DNLSE, as in Eq.~\eq{e3}, approaches the quantum
results, as per the TWA. For instance, if the ratio of non-condensate particles
to condensate particles in the initial state is large, one may expect the classical DNLSE
to provide an especially poor approximation.  Now, the expression~\eq{ee19} for the amplitudes
$v_q$ shows that the interaction strength enters through the dimensionless parameter
$\chi/J$. By virtue of our normalizations, the interaction parameter $\chi$, Eq.~\eq{CHIDEF},
is proportional to the number of atoms. If the number of atoms is increased while adjusting
the strength of the atom-atom interactions in such a way that $\chi/J$ is held constant,
the number of noncondensate atoms remains constant and the ratio of non-condensate
particles to condensate particles in the initial state is reduced. We expect the error in DNLSE
correspondingly become smaller~\cite{RUO05}. In fact, even if the non-condensate population
in the initial state vanishes, the relative contribution of the vacuum Wigner noise is reduced
when the condensate population is increased, and we again expect the error in DNLSE to
be reduced.

The potential instability, however, adds a new element to the picture. Namely, for the consistency of our present development the excitation modes must be normalized to $|u_q|^2-|v_q|^2=1$, whereas for an unstable mode necessarily $|u_q|^2-|v_q|^2=0$ holds true. This means that when the point of dynamical instability is approached continuously, for some mode $q$ the amplitudes diverge, $|u_q|\rightarrow\infty$ {\em and\/} $|v_q|\rightarrow \infty$. By Eq.~\eq{V15} the number of noncondensate atoms also diverges, $\bar{N}_n\rightarrow\infty$. At the point of instability steady-state quantum fluctuation diverge, according to the Bogoliubov approximation. The Bogoliubov method is based on the assumption of {\em small\/} fluctuations so that it in itself fails for an unstable system, but it is clear that the onset of instability is associated with large quantum fluctuations.

\begin{figure}
\includegraphics[clip,width=1.0\linewidth]{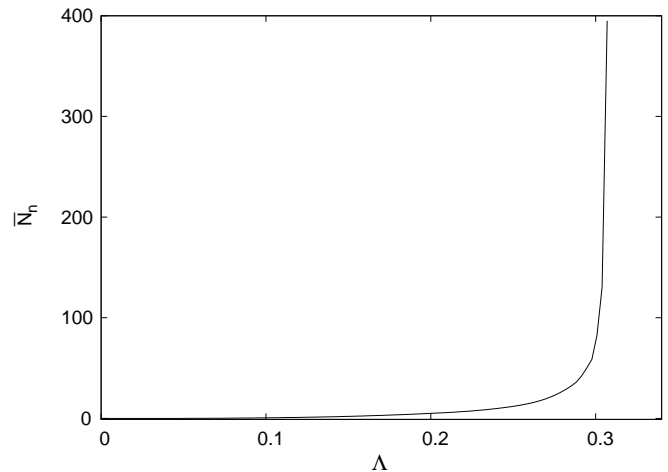}
\caption{Number of noncondensate atoms $\bar{N}_n$ given by Eq.~(\ref{V15}) as a function of the scaled interaction strength $\Lambda$ within Bogoliubov approximation for $L=32$. Note that the critical interaction strength at the onset of the instability is 0.308.}
\label{fp10}
\end{figure}

While the asymptotic results are easy to infer, only a detailed analysis can tell what the number of noncondensate atoms is in a specific situation. Figure~\ref{fp10} shows the number of noncondensate atoms $N_n$ as a function of the interaction parameter $\Lambda$ for the number of lattice sites $L=32$. Both the generally very small number of noncondensate atoms for weak interactions and the divergence at the onset of instability  $\Lambda  = 0.308 \simeq \pi^2/L$ are evident.

\subsection{\label{sec:level3ins}Numerical realization}

We study the non-equilibrium quantum dynamics of a BEC within TWA. We consider a BEC in a lattice that is initially in a stable steady flow state. This is crucial for the validity of the Bogoliubov approximation for the initial \colr{state as well as for the generation of the stochastic initial noise in the TWA}. A stable state of the atomic gas can be experimentally driven to a dynamically unstable state, for instance, by starting with a moving lattice with the quasimomentum of the atoms $p<\pi/2$ and accelerating the lattice so that the momentum satisfies $p > \pi/2$. In our examples, though, we envisage suddenly modifying atom-atom interactions. Henceforth we always have the number of lattice sites $L=32$ and the initial flow state with $p=\pi$. We perform the simulations for two different types of initial states: noninteracting atoms with all atoms in the condensate, and an interacting state with a nonzero condensate depletion. As to the interacting initial state, we select the dimensionless interaction strength $\Lambda = \chi/2J=0.284$. For this value the average noncondensate atom number is $\bar N_{n}\simeq 30$ whereas the critical value of $\Lambda$ at the onset of the instability is 0.308 , as per Fig.~\ref{fp10}.  In all cases considered here, the initial depletion $\colr{\bar N_n}/N$ is less than 10\%, and the Bogoliubov approximation for the initial state  is presumed accurate.  At the beginning of the time evolution the atom-atom interactions are turned up instantaneously to the dynamically unstable value of $\Lambda=0.48$. We investigate the effect of quantum fluctuations by varying the total atom number $N$ and the atomic interaction strength $g$ (Eq.~\eq{CHIDEF}) so that the interaction parameters $\chi$ and $\Lambda$ (and the chemical potential) remain constant.

For each individual realization of the time evolution of the ensemble of the Wigner distributed  wave functions we sample the initial state as explained in the previous Section~\ref{sec:level3ins}. The random numbers $(\alpha_q,\alpha^*_q)$ are obtained using the Box-Mueller algorithm \cite{NR}. As before, we integrate the dynamical equations~(\ref{V1}) using the FFT split-step method \cite{Javanainen}. Depending on the task at hand, we either study the individual trajectories $\psi_\idx^W(t)$, or averages over a number of trajectories.

It should be noted that our process contains an uncontrolled approximation. There are some \colr{indications~\cite{ISE06,RUO05,BER07,KOR08}} that time evolution as prescribed by TWA could becomes exact in the asymptotic limit $N\rightarrow\infty$ with a fixed $\chi$, but even if this were the case, the asymptotic limit per se does not tell how good the results are for given values of the parameters. For instance, we do know how long the TWA time integration remains valid. This question should be studied fully quantum mechanically, but at the present time we do not know of any framework to address the issue. How would one describe the modulational instability of a classical nonlinear system ab initio using  linear quantum mechanics, which by definition prescribes a stable evolution of any initial state? Moreover, in a classical instability the translation invariance of the state of the lattice is broken, but unitary time evolution from quantum mechanics will not spontaneously break any symmetry of the Hamiltonian. For the time being we rely on the practical observation from working with the TWA method that it usually signals its own demise. Typically, upon integrating long enough, each individual trajectory seemingly loses any relation to the initial state. In our computations we have not encountered a situation of this kind.

\subsection{Results}

Since the Wigner functions govern symmetrically ordered expectation values, we need certain transformations to calculate the ordinary normally ordered expectation values from the simulation data \cite{ISE05,ISE06}. Here we only consider the lowest-energy band in the tight-binding approximation, so normally ordering the operator expectation values is straightforward. According to Ref.~\cite{ISE06}, we have   the atom number in a lattice site
\begin{equation}
n_\idx=\langle\psi^*_\idx\psi_\idx\rangle_W -\frac{1}{2}\,,
\end{equation}
with the corresponding fluctuations
\begin{equation}
\Delta n_\idx=\sqrt{\langle(\psi^*_\idx\psi_\idx)^2\rangle_W-\langle\psi^*_\idx\psi_\idx\rangle_W^2-\frac{1}{4}}\,.
\end{equation}
The overlap of the field amplitudes $\psi_l(0)$ and $\psi_l(t)$, which is a measure of the revival of the pulse, is given by
\begin{equation}
f^W(t)= \langle\left| \hbox{$\sum_\idx$}\psi_\idx^*(t)\psi_\idx(0)\right|^2\rangle_W.
\end{equation}

\begin{figure}
\includegraphics[clip,width=1.0\linewidth]{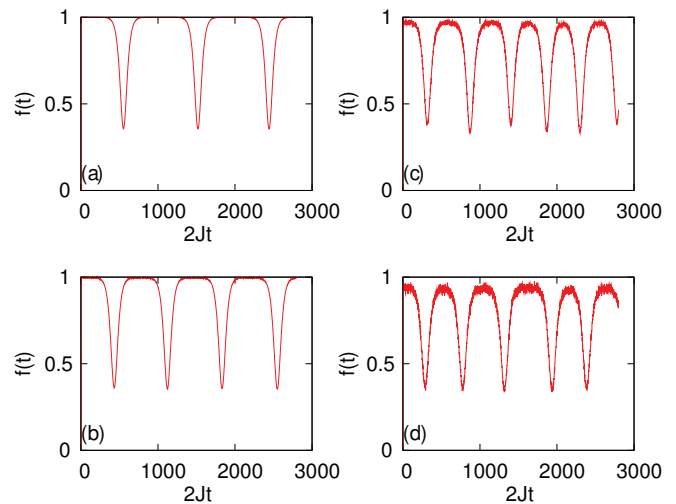}
\caption{(Color online) Overlap of a single stochastic realization with the initial state as a function of time sampled from the Wigner distribution for a fixed chemical potential with different atom numbers (a) $N=10^6$, (b) $N=10^4$, (c) $N=10^3$ and (d) $N=500$. The initial state is interacting with  $\bar N_{n}=30$. The parameters of the simulation are $\Lambda=0.48$, $L=32$, and $p=\pi$. There is no significant damping in the oscillation even if the noncondensate noise is substantial.}
\label{f10}
\end{figure}

In Fig.~\ref{f10} we show typical single-trajectory results for the overlaps $f(t)$ for different values of the atom-atom interactions $g$ and the total atom number $N$. All different cases have the initial interaction parameter $\Lambda=Ng/(2J)=0.284$ that is instantaneously changed to $\Lambda=0.48$, but the values of $N$ and $g$ are varied individually so that the total number of atoms is (a) $N=10^6$, (b) $N=10^4$, (c) $N=10^3$, and (d) $N=500$.

The number of noncondensate atoms in the Bogoliubov approximation does not change whenever we vary $\bar N_c$ and $g$ for a fixed value of $\bar N_c g$. For a smaller total number of atoms $N$, the noncondensate atom fraction $\bar N_{n}/N$ in the initial state and the coupling constant $g$ are larger and, consequently, quantum effects are stronger. Each plot in Fig.~\ref{f10} clearly shows the pulsating instability without any noticeable damping. Quantum fluctuations and the initial noncondensate population act as a seed for the scattering processes taking atoms out of the ground state. The period of the pulsating instability depends on the interaction coefficient $g$, albeit not dramatically; the larger the value of $g$ (corresponding to the smaller value of $N$), the shorter the period of oscillation, since higher scattering rate leads to a faster ground state depletion and, consequently, a shorter oscillation period.

\begin{figure}
\includegraphics[clip,width=1.0\linewidth]{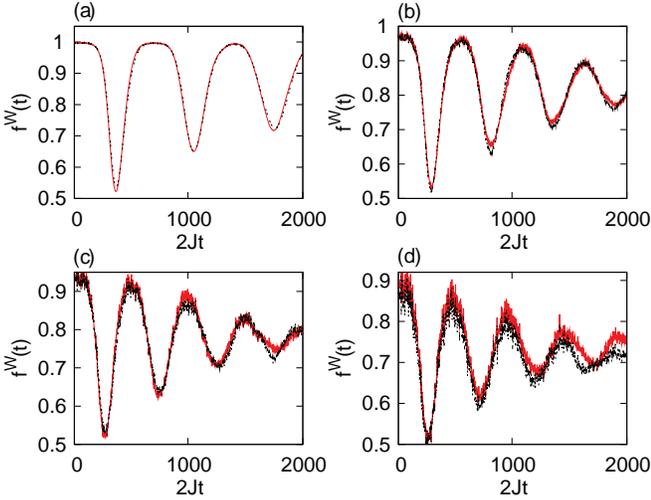}
\caption{(Color online) Comparison of the ensemble average of the overlap of the state of the lattice sampled over $400$ realizations for the initial number of non-condensate atoms $\bar N_n=0$ (solid line) and $\bar N_n=30$ (dashed line), and for the total number of atoms (a) $N=10^4$, (b) $N=10^3$, (c) $N=500$ and (d) $N=300$. The parameters of the simulations are $\Lambda=0.48$, $L=32$, and $p=\pi$. The two different initial states represent similar time evolution when the total atom number is large. However, the curves start deviating as the number of atoms is reduced.}
\label{f21}
\end{figure}

Figure \ref{f21} represents an ensemble average of the overlap $f^W(t)$ sampled over 400 trajectories for the total number of atoms (a) $N=10^4$, (b) $N=10^3$, (c) $N=500$ and (d) $N=300$ for the interacting (solid curve) and the noninteracting (dashed curve) initial states. In all cases the final value of the interaction parameter is again $\Lambda=0.48\,\,(\propto N g)$, but the nonlinearity $g$ and the total atom number $N$ are different.

\begin{figure}
\includegraphics[clip,width=1.0\linewidth]{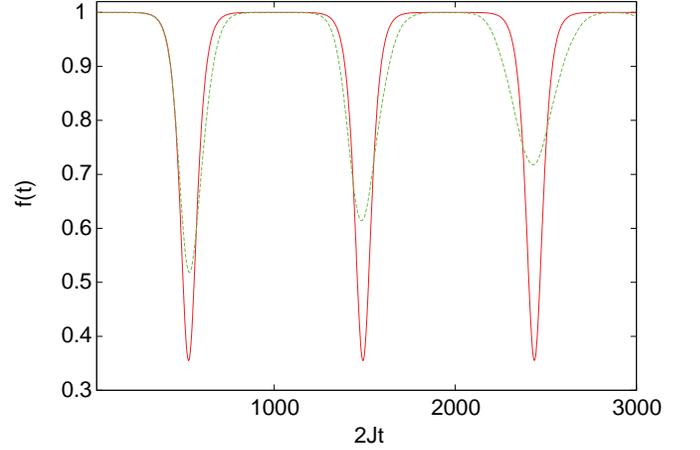}
\caption{(Color online) Comparison of the overlaps in a typical single realization (solid line) versus an ensemble average sampled over 400 realizations (dashed line) with the initial interacting state for $\bar N_{n}=30$ and $N=10^6$. The parameters of the simulations are $\Lambda=0.48$,  $L=32$ and $p=\pi$.}
\label{f12}
\end{figure}

While the almost undamped quasiperiodic behavior is seen in individual stochastic realizations,
the quantum mechanical ensemble averages of the wave function revival become progressively weaker when the effective interaction strength is increased. This is because the shape and the timing of the pulsations in each realization change due to quantum fluctuations. Enhanced quantum fluctuations for smaller $N$ (and larger $g$) are clearly observable in increased damping rates. For the interacting initial state, the initial noncondensate atom number $\bar N_{n}=30$ provides a nonnegligible noncondensate atom fraction only if the total atom number is small, and so the results for the noninteracting and interacting initial states differ significantly only for small total atom numbers. Figure~\ref{f12}  represents a comparison of the overlaps in a typical single realization and in an ensemble average for the large atom number $N=10^6$, and with the interacting initial state. The other parameters of the simulations are same as in Fig.~\ref{f21}.

\begin{figure}
\includegraphics[clip,width=1.0\linewidth]{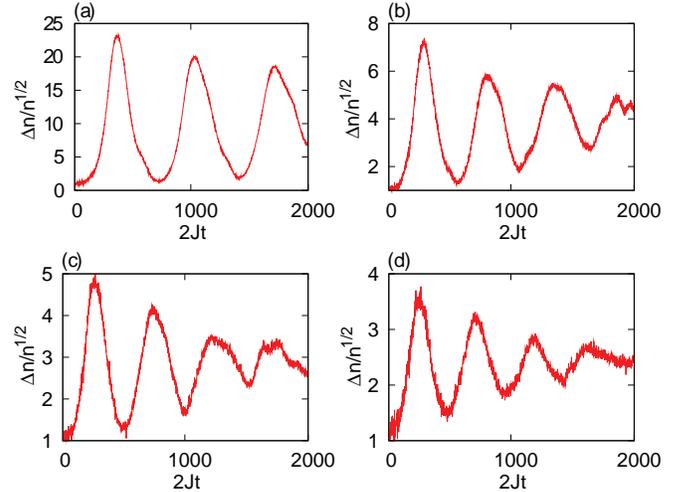}
\caption{(Color online) Atom number fluctuations  in the central lattice site as a function of time sampled over $400$ realizations for the total numbers of particles (a) $N=10^4$, (b) $N=10^3$, (c) $N=500$ and (d) $N=300$ with $N_{n}=30$. We use the interacting initial state. }
\label{f13}
\end{figure}

Similar damping is also observed in the atom number fluctuations at a fixed site, say, center of the lattice. In Fig.~\ref{f13} we show ensemble averages of the atom number fluctuations $\Delta n/\sqrt{n}$ for the total number of the atoms (a) $N=10^4$, (b) $N=10^3$, (c) $N=500$ and (d) $N=300$ and for the interacting initial state. The parameters of the simulations are as before. The pulsating instability causes strongly super-Poissonian atom number fluctuations at any fixed lattice site. 

The main focus in this work is to estimate the inherent quantum effects on the pulsation phenomenon of the BEC in an optical lattice. We have found that quantum fluctuations have an effect on the collapse and revival of the pulse. In the case of a single realization the revival is very robust and repeats over a long time. The ensemble averages of the overlap of the wave function revival display a recognizable damping behavior due to quantum fluctuations. We have also found that the damping rate is increased if the initial state is depleted with a nonnegligible noncondensate fraction.

\begin{figure}
\includegraphics[clip,width=1.0\linewidth]{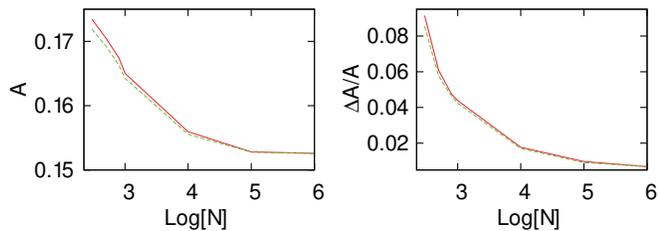}
\caption{(Color online) Ensemble average of the amplitude (left) of the first peak and its fluctuations (right) during pulsation as a function of the total number of atoms, $N$, for a fixed chemical potential for interacting (solid line) and non-interacting (dashed line) initial states. The parameters of the simulation are same as in Fig.~\ref{f21}. The amplitude reaches the classical value for higher atom numbers. Quantum noise increases the amplitude and the fluctuations of the amplitude for small atom numbers.}
\label{f14}
\end{figure}

The TWA also provides for computing of quantities that are affected by quantum fluctuations and could be readily measurable, but may be very difficult to extract from the full quantum solution of the problem if it were on hand. Parameters for the shape of the pulse, obscured in quantum mechanics by averaging over the random time and position of the pulse, are an example. In Fig.~\ref{f14} we show the amplitude for the first pulse in the pulsating instability as a function of the total number of atoms (for a fixed nonlinearity, as before), and its fluctuations.
These are averaged over 400 realizations, both for the interacting (solid curve) and the noninteracting (dashed curve) initial states. The parameters of the simulations are the same as in Fig.~\ref{f21}. 

The pulse amplitude approaches a classical value that may be obtained from the GPE in the limit of weak quantum fluctuations (here corresponding to large $N$ and small $g$). However, both the amplitude of the atom pulse and the amplitude fluctuations grow when the atom number becomes small and quantum fluctuations become more dominant. Similarly to the case of the overlap revival, the distinction between the two initial states is apparent only when the noncondensate fraction is nonnegligible.

\section{Concluding Remarks}

In this paper we have presented new insights into the unstable dynamics of the BEC in an optical lattice in the limit of weak atom-atom interactions, with quantum fluctuations included. The common belief is that the flow of the dynamically unstable BEC in an optical lattice would be erratic, or lead to the formation of stable solitons. Here we move a step further and show that, in classical mean-field theory, the instability may also trigger a quasi-periodic pulsation in the atom density distribution if the atom-atom interactions are weak. The requirement that linear stability analysis finds a single unstable mode gives the scale for the  `weak' nonlinearity and the ensuing pulsating phenomena.

A qualitative argument has been put forward to explain the pulsating behavior of the dynamics by comparing the lattice system with the integrable double-well system. In the case of two wells the unstable mode leads to a non-trivial dynamics in the population imbalance such that an infinitesimal noise could produce a large-amplitude collective oscillation of the atoms between the wells. We surmise that the pulsating instability is a remnant of the integrability as manifest in the two-well system.

We incorporate quantum fluctuations using stochastic phase-space methods. We use the Bogoliubov approximation to generate the initial state for the time evolution of the system. A sequence of the stochastic fields obtained in this way are then used to calculate expectation values of the observables. We also compare the single realization results with the ensemble averages. Generally speaking, the pulsating behavior survives in the face of quantum fluctuations. It is observed that the quasiperiodic behavior in the time dynamics can still be seen in single realizations. However, the quantum averages show that the revivals of the pulses tend to get washed out as the atom number gets smaller \colr{while the chemical potential is held constant}.

For experimental realizations, the flow states $p\approx \pi$ near the Brillouin zone boundary can be prepared by accelerating the lattice \cite{FAL04}. Alternatively, by exploiting the symmetry of the DNLSE, for every solution $\psi_\idx(t)$ with the given interaction parameter $\Lambda$ there is a solution $(-1)^\idx\psi_\idx ^*(t)$ for $-\Lambda$. That means the state for $p=\pi$ in the repulsive case is equivalent to the state for $p=0$ in the attractive case. Manipulation of the sign of the interactions gives much additional leeway for the experiments. However, given that the pulsating phenomenon only occurs in the limit of weak nonlinearity, the corresponding time scales for the pulsation can be very long and pose severe technical challenges.

\colr{\acknowledgements{We acknowledge financial support from NSF (PHY-0750668) and EPSRC.}}

\end{document}